\documentclass[12pt]{article}
\usepackage{amsmath}
\usepackage{amssymb}
\usepackage{amscd}
\usepackage{tabularx}
\usepackage{float}
\usepackage{theorem}
\usepackage{enumerate}
\def\eqnum#1{}

\begin{document}
\title {Interevent time distribution in seismicity: \\
a theoretical approach}
\author
{
G. Molchan  \thanks{E.mail adress: molchan@mitp.ru} \\
International Institute of Earthquake Prediction Theory \\
and Mathematical Geophysics, \\
Warshavskoe shosse 79, k.2, Moscow 117556 \\
and \\
The  Abdus Salam International Centre for Theoretical Physics, \\
SAND group, Trieste, Italy
}
\date{}
\maketitle
\begin{abstract}
This paper presents an analysis of the distribution of the time
$\tau$  between two consecutive events in a stationary point
process. The study is motivated by the discovery of a unified
scaling law for  $\tau$  for the case of seismic events. It is shown
that the unified law must necessarily be exponential. We discuss
the parameterization of the empirical unified law and the
physical meaning of the parameters involved. \\

\noindent PAC numbers: 91.30.Dk,05.65.+b,89.75.Da
\end{abstract}

\newpage

\section{Introduction}

The studies \cite{Bak},\cite{Cor},\cite{Corr} have
discovered a new scaling law for
seismic events on the phase space location-time-magnitude.
Specifically, the distribution density for time $\tau$ between two
consecutive events of magnitude $m>m_c$ has the form

\begin{eqnarray}
p_\tau(t) =\lambda f(\lambda t)
\label{eqc1}
\end{eqnarray}
where $\lambda$ is the rate of events with $m>m_c$ in a given area $G$,
while $f$ is a universal function that is independent of the choice
of $G$ and cutoff magnitude $m_c$. The relation (\ref{eqc1}) is astonishing,
being tested (as it has been in \cite{Corr}) for a very wide range of $m_c$
(between 2 and 7.5), for seismic regions $G$ of very different
linear size $L$ (between 20 km and the size of the Earth), as well
as for different catalogs, both regional and global ones, and
different time periods.

The parameterization of $f$ seems not yet to have settled
down. According to \cite{Corr}:

\begin{eqnarray}
f(x) = cx^{\gamma  -1}\exp(-x/a)
\label{eqc2}
\end{eqnarray}
in the region $x\ge 0.05$ with $\gamma =0.74\pm 0.05$ and $a=1.23\pm .15$.
The pioneering work \cite{Bak} uses a parameterization of (\ref{eqc2}) for
the whole range of $x$ with $\gamma =0.1$
(see refined estimates in \cite{Cor}). This allows the behavior of $f(x)$
about zero to be interpreted in terms of the Omori law.

The subsequent discussion strives to answer the following
questions:

What is the distribution of $\tau$ in examples of synthetic
seismicity?

What is the physical meaning of the parameters $\gamma$ and $a$ in
(\ref{eqc2})?

Assuming the form of $f$ to be universal for $\tau$, what should it
be?

\bigskip
\bigskip

\section{The Model}

Earthquakes frequently form anomalous clusters in space-time.
The largest event in a cluster is termed the {\it main} event.
The events that occurred before and after the main event in a
cluster are called fore- and aftershocks, respectively. It is
assumed in a zero approximation that main events constitute a
time-uniform Poisson process. That assumption is widely
employed in seismic risk studies.

Aftershocks dominate clusters both as regards their
number and duration. Their rate as a function of time is
described by the Omori law:

\begin{eqnarray}
n(t) = ct^{-p}, \quad t>t_0,
\label{eqc3}
\end{eqnarray}
where $t_0$ is small. Relation (\ref{eqc3}) holds fairly well during the
first few tens of days (up to a year) with the parameter $p=0.7-1.4$
\cite{Uts}. At large times the value of p becomes greater, occasionally
significantly so, making $n(t)$ decay in an exponential manner.
Taken on the whole, background seismicity and spatial
interaction do not allow reliable conclusions to be drawn for the
Omori law at large times. Cases in which (\ref{eqc3}) holds during
decades are unique \cite{Uts}.

Following the above description, we consider the
following model for seismic events in time. The spatial and
magnitude components of an event are disregarded for
simplicity of reasoning. Let $\{x_i\}$ be a homogeneous point
Poisson process on a line with rate $\lambda^*$. It is an analogue of
main events. Let $N_0(dt)$ be an inhomogeneous point process with
rate $\delta(t)+\lambda_0(t)$. Here, $\delta$ is the delta function,
while the presence of $\delta (t)$ means that the event $t=0$ belongs
to $N_0$. The notation $N_0(\Delta)$ defines the number of
events $N_0$ in the interval $\Delta$. We will assume that

\begin{eqnarray}
\int \lambda_0(t)\,dt = \Lambda < \infty.
\label{eqc4}
\end{eqnarray}
This requirement ensures that the total number of events in
$N_0$ is a.s. bounded.

Consider the infinite series $N_0^{(i)}(dt)$, $i=0,\pm 1,\pm 2,\ldots$
of independent samples of $N_0$. The theoretical process $N$ is the sum

\begin{eqnarray}
N(dt) = \sum_iN_0^{(i)}(dt-x_i).
\nonumber
\end{eqnarray}
The process $N_0^{(i)}$ that has been shifted by the amount $x_i$
can be associated with the cluster of the main event $x_i$.

Our task is to describe the distribution of $\tau$ between
two consecutive events in $N$. The distribution is uniquely
specified, because the process $N$ is stationary. It is also easy to
see that the rate of $N$ is

\begin{eqnarray}
\lambda = \lambda^*(1 + \Lambda).
\nonumber
\end{eqnarray}
According to \cite{Dal},

\begin{eqnarray}
P(\tau > t) = \frac{\partial}{\lambda \partial a}\,P\{N([a,t]) = 0\}\,
\bigg \vert_{a=0}
\label{eqc5}
\end{eqnarray}
and

\begin{eqnarray}
P(N(\Delta) = 0) = \exp \{-\lambda^*\int P(N_0(\Delta -x) > 0)\,dx\}.
\nonumber
\end{eqnarray}
The last relation follows from the fact that the main events are
poissonian. Since the aftershocks make the bulk of a cluster, we
shall assume in what follows that $\lambda_0(t)=0$ for $t>0$. Consequently,

\begin{eqnarray}
P(N_0(\Delta) > 0) = \left\{\begin{array}{ll}
1, \quad \mbox {\rm if} \quad 0 \in \Delta \\
0, \quad \mbox {\rm if} \quad \Delta \subset (-\infty, 0).
\end{array}\right.
\nonumber
\end{eqnarray}
Combining the above relations, one gets

\begin{eqnarray}
P(\tau > t)&=&\exp \{-\lambda^* \int\limits_0\limits^{\infty}
P(N_0(u, t+u)>0)\,du - \lambda^*t\} \times \\
\nonumber
&\times& [P(N_0(0,t)=0) + \int\limits_0\limits^{\infty}
P(N_0(du)>0, \,N_0(u, u+t) = 0)] / (1+\Lambda).
\label{eqc6}
\end{eqnarray}
We now describe the behavior of the distribution of $\tau$ near 0
and $\infty$.

{\bf Statement 1.} (a) {\it If cluster duration has a finite
mean}, $\bar{\tau}_{cl}$, {\it then}

\begin{eqnarray}
P(\tau >t) = \exp \,(-\lambda^*
(t+\bar{\tau}_{cl})) /(1+\Lambda) \cdot
(1+o(1)), \quad  t \to \infty.
\nonumber
\end{eqnarray}

(b) {\it Let}
$\lambda_0(t) \sim ct^{-1-\theta}$,\,  $t\to \infty$
{\it where} $0<\theta <1$. {\it Then}

\begin{eqnarray}
P(\tau >t) = \exp \,(-\lambda^*t-O(t^{1-\theta})) / (1 + \Lambda), \quad
t \to \infty.
\label{eqc7}
\end{eqnarray}
In other words, one has

\begin{eqnarray}
\lim\limits_{t\to \infty}\ln P(\tau >t) /(\lambda t) = \lambda^*/\lambda
\nonumber
\end{eqnarray}
for a Poisson sequence of main events in a broad class of cluster
models. In terms of the parameterization of (\ref{eqc2}), that means that

\begin{eqnarray}
a = \lambda /\lambda^* = 1 + \Lambda.
\nonumber
\end{eqnarray}
With $a=1.23$ (as in \cite{Corr}) the main events make $a^{-1}\simeq 81\%$ of
the total number of events.

The following regularity conditions should be imposed
on $N_0$ in order to be able to describe how the distribution
density for $\tau$ behaves for small $t$:

\begin{eqnarray}
P(N_0(u, u+t) > 0& \vert &N_0\{ u+t\} =1) = o(1), \qquad \qquad \,\,t \to 0
\label{eqc8}
\end{eqnarray}
\begin{eqnarray}
\quad
P(N_0(u, u+t) > 0&\vert &N_0\{ u\} =1, N_0\{ u+t\} ) = o(1), \quad t \to 0
\quad
\label{eqc9}
\end{eqnarray}
where the notation $\vert $ denotes conditional probability, and
$N_0\{ s\} =1$ means that there is an event at the point $s$. We assume in
addition that (\ref{eqc8}), (\ref{eqc9}) hold uniformly in $u\ge 0$.

That last requirement is no limitation for the case of
seismic events, considering that the rate of cluster events and
time relations between them seem to be rapidly decaying over
time. The requirements (\ref{eqc8}), (\ref{eqc9}) themselves ensure
that two very
closely lying cluster events are not likely to contain another
cluster event between them, that is, (\ref{eqc8}), (\ref{eqc9})
express the requirement
of sparseness or repulsion for events that are close in time.
It follows from the obvious inequality
$P(N_0(\Delta)>0)<EN_0(\Delta)$
that (\ref{eqc8}), (\ref{eqc9}) will hold, if one demands that

\begin{eqnarray}
E(N_0(u, u+t) \vert {\cal A}) = o(1), \quad t \to 0,
\nonumber
\end{eqnarray}
where ${\cal A}=(N_0\{ u+t\} =1)$ in the case (\ref{eqc8}) and
${\cal A}=(N_0\{ u\} =1, N_0\{ u+t\} =1)$ in the case (\ref{eqc9}).

{\bf Statement 2.} {\it If} (\ref{eqc8}), (\ref{eqc9}) {\it hold, the probability density for}
$\tau$ {\it (provided it exists) has the following form as} $t\to 0$:

\begin{eqnarray}
p_\tau (t) = [\lambda_0(t) + \int\limits_0^\infty
\lambda_0(u)\lambda_u(t)\,du + \lambda (1+\Lambda)] /
(1+\Lambda) \cdot (1+o(1)),
\label{eqc10}
\end{eqnarray}
{\it where}
$\lambda_u(t) =P(N_0(t+u,\, t+u+\delta)>0\, \vert \,N_0\{ u\} =1)/\delta$
{\it is the conditional rate of} $N_0$ {\it after time u given a cluster
event has occurred at that time.
In particular, if} $\lambda_0(t)\uparrow \infty$ as $t\to 0$ {\it and}

\begin{eqnarray}
\lambda_u(t) < k\,\lambda_0(t), \quad 0<t<\varepsilon,
\label{eqc11}
\end{eqnarray}
{\it then}

\begin{eqnarray}
1<p_\tau(t) / \lambda_0(t)<c, \quad t\to 0.
\nonumber
\end{eqnarray}
In other words, when (\ref{eqc8}), (\ref{eqc9}) hold, the distribution density for $\tau$
for small t is proportional to the rate of cluster events
immediately after the main event. The statement is not obvious,
since any interevent interval is not necessarily started by a main
event.

The proofs of the Statements have been relegated to the Appendix.

\bigskip
\bigskip

\section{Examples}

Examples will now be discussed to be able to understand how
far the above assumptions are restrictive.

{\it The trigger model}. Historically, this is the first seismicity model
to appear (see \cite{Ver}). It assumes the cluster process $N_0$ to be
poissonian. The model has not found acceptance in seismicity
statistics, because the likelihood of an observed sample in that
model is technically difficult to use. This does not rule out that
the model may be helpful, however.

Because increments in $N_0$ are independent, the
requirement (\ref{eqc8}), (\ref{eqc9}) has the form

\begin{eqnarray}
P(N_0(u, u+t) > 0) = \int\limits_u\limits^{u+t}\lambda_0(x)\,dx
= o(1), \quad t \to 0.
\nonumber
\end{eqnarray}
If $\lambda_0(x)$ is a decreasing function, one has

\begin{eqnarray}
\int\limits_u\limits^{u+t}\lambda_0(x)\,dx <
\int\limits_0\limits^t\lambda_0(v)\,dv = o(1).
\nonumber
\end{eqnarray}
Consequently, the
decrease of $\lambda_0(x)$  ensures that (\ref{eqc8}), (\ref{eqc9}) take
place uniformly in $u$. The same property of $\lambda_0(x)$  also
ensures (\ref{eqc11}):

\begin{eqnarray}
\lambda_u(t) = \lambda_0(u+t) < \lambda_0(t).
\nonumber
\end{eqnarray}
We now are going to refine the asymptotic form of $p_\tau(t)$ for
small $t$.

{\it Let} $\lambda_0(x)$ {\it be a smooth decreasing function and}
$\lambda_0(t)=ct^{-p}$,  $t<1$.
{\it Then}

\begin{eqnarray}
p_\tau (t) \simeq ct^{-p} + c_1t^{-\alpha} +c_2, \quad t \to 0,
\nonumber
\end{eqnarray}
{\it where} $\alpha =2p-1$ {\it for} $p>1/2$ {\it and}
$\alpha=0$ {\it for} $p\le 1/2$.

This can be seen as follows. When $p>1/2$, one has

\begin{eqnarray}
\nonumber
I_t&=&\int\limits_0\limits^\infty \lambda_0(u) \lambda_u(t) =
c^2\int\limits_0\limits^1u^{-p}(u+t)^{-p}\,du +
\int\limits_1\limits^\infty \lambda_0(u) \lambda_0(u+t)\,du \\
\nonumber
&=&c^2t^{1-2p}\int\limits_0\limits^\infty u^{-p}(1-u)^{-p}\,du +
\mbox {\rm const} + o(1), \quad t\to 0.
\nonumber
\end{eqnarray}
When $p<1/2$, one has

\begin{eqnarray}
I_t=\int\limits_0\limits^\infty \lambda_0^2(u)\,du + o(1).
\nonumber
\end{eqnarray}

{\it The self-exciting model}.
A cluster in this model is generated by the
following cascade process. The first event $t=0$ is defined as the
event of rank 0. It generates a Poisson process with rate $\pi_0(t)$;
its events $\{ t_i^{(1)}\}$ are ascribed rank 1. The procedure then
becomes recursive: each event $\{ t_i^{(r)}\}$ of rank $r=1,2,...$ generates
a Poisson process of its own which is independent of the
previous ones and which has the rate
$\pi_0(t-t_i^{(r)})$. The offspring of a
rank $r$ event are events of rank $r+1$, the events of all ranks
constituting the desired cluster  $N_0$.

The process $N$ with clusters as described above is known
as the self-exciting model \cite{Haw} or the epidemic type model \cite{Oga}.
The model is rather popular in the statistical studies and
forecasting of seismicity thanks to the fact that the predictable
component of $N$ has simple structure:

\begin{eqnarray}
E(N(t+\delta) - N(t) > 0 \,\vert \,{\cal A}_t) =
\sum_{t_1<t} \pi_0(t-t_i) \cdot \delta +\lambda^*\delta,
\nonumber
\end{eqnarray}
where the $t_i$ are events of $N(dt)$ and
${\cal A}_t=\{ N(ds), s<t\}$ is a past of the process.

It is easy to see that the rate $\lambda$ of the process $N$ is
bounded, if

\begin{eqnarray}
\lambda_{\pi} = \int\limits_0\limits^\infty \pi_0(t)\,dt < 1,
\nonumber
\end{eqnarray}
also,
\begin{eqnarray}
\Lambda = \lambda_{\pi}/(1-\lambda_{\pi}) \quad \mbox {\rm and}
\quad  \lambda = \lambda^*/(1-\lambda_{\pi}).
\nonumber
\end{eqnarray}

{\bf Statement 3.} (a) {\it The cluster rate function for the self-exciting
model is}

\begin{eqnarray}
\lambda_0(t) = \pi_0(t) + \pi_0 * \pi_0(t) + \pi_0 * \pi_0 * \pi_0(t) + \ldots,
\quad t>0,
\label{eqc12}
\end{eqnarray}
{\it where} $*$ {\it denotes the convolution}.

{\it Let} $\pi_0(t)$ {\it be monotone near} 0, {\it where}
$\pi_0(t)\sim At^{-p}$, $0<p<1$. {\it Then}

\begin{eqnarray}
\lambda_0(t) / \pi_0(t) \sim 1, \quad t\to 0.
\nonumber
\end{eqnarray}
{\it Let} $\pi_0(t)$ {\it be monotone at} $\infty$,
{\it where} $\pi_0(t)\sim Bt^{-1-\theta}$, $0<\theta <1$. {\it Then}

\begin{eqnarray}
\lambda_0(t) / \pi_0(t) \sim (1-\lambda_{\pi})^{-2}, \quad t\to \infty.
\nonumber
\end{eqnarray}

(b) {\it The distribution density for} $\tau$ {\it as} $t\to 0$
{\it has the form}

\begin{eqnarray}
p_\tau(t) = [(1-\lambda_{\pi})\lambda_0(t) +
\int\limits_0\limits^\infty \lambda_0(x)\lambda_0(x+t)\,dx + \lambda]
\cdot (1+o(1)), \, \, t\to 0.
\label{eqc13}
\end{eqnarray}
{\it Let} $\pi_0(t)$
{\it be monotone near zero, where} $\pi (t)\sim At^{-p}$,  $0<p<1$;
{\it let} $\pi_0(t)<\varphi (t)$,
{\it where} $\varphi$ {\it is a smooth function},
$\int\limits_0\limits^\infty \varphi (t)\,dt<1$,
$\varphi(t) \sim ct^{-1-\theta}$,\, $t\to \infty$,\, $0<\theta <1$.
{\it Then}

\begin{eqnarray}
p_\tau(t) = O(\lambda_0(t)) \quad \mbox {\rm as} \quad t\downarrow 0.
\nonumber
\end{eqnarray}

{\it The time-magnitude self-exciting model}.
The self-exciting model is frequently considered
on the time-magnitude space as follows (see, e.g., \cite{Sai}): each event
$t_i$ (both when a main or a cluster one) is ascribed a random
magnitude $m_i$. The $m_i$ are independent for different
$t_i$ and have identcal distributions with density $p(m)$. The
generation of clusters is that described above, the only
difference being that an event $(s,m)$ generates a cluster with rate
$q(m)\pi (t-s)$. It can be assumed without loss of generality that
$\int q(m)p(m)\,dm = 1$.
This normalization preserves statements 1, 3 for the self-exciting
process $(t,m)$ as well, independent of the choice of $p(m)$ and
$q(m)$. The function $\lambda_0(t)$ as given by (\ref{eqc12})
then corresponds to
the cluster rate when averaged over magnitude $m$. For purposes
of seismology, $p(m)$ corresponds to the normalized Gutenberg-
Richter law, $p(m)=\beta e^{-\beta (m-m_0)}$,\,$m>m_0$
while $q(m)=e^{\alpha (m-m_0)}(1-\alpha/\beta)$
is proportional to the size of the
cluster that has been triggered by an event of magnitude $m$.

\bigskip
\bigskip

\section{The unified scaling law}

According to \cite{Corr}, the distribution of $\tau$ depends on
the single parameter $\lambda$, see (\ref{eqc1}). The parameter $\lambda$ is
specified by the choice of the area and cutoff magnitude $m_c$. This
allows variation of $\lambda$ in a very wide range. Experiments
which test (\ref{eqc1}) in \cite{Corr} concern both the Earth as a whole and
smaller or larger areas of it. One can always select such areas in
which seismicity is weakly interdependent. For the theoretical
analysis of the unified scaling law (\ref{eqc1}) one may be interested in
the following

{\bf Statement 4.} {\it Assume that it is possible to choose two
regions} $G_1$ {\it and} $G_2$ {\it with independent stationary
sequences of events} $N_i(dt)$.
{\it If the unified scaling law} (\ref{eqc1}) {\it holds for} $G_1$, $G_2$
{\it and} $G_1\cup G_2$ {\it and} $f(x)<cx^{-\theta}$,\,
$0<\theta <1$ {\it for small} $x$,
{\it then} $f(x)=\exp (-x)$.

{\it Proof}. By (\ref{eqc5}),

\begin{eqnarray}
p_\tau(t) =
\frac{\partial^2}{\lambda \partial t^2}\,P\{N(0,t) = 0\},
\label{eqc14}
\end{eqnarray}
where $\lambda$ is the rate of $N(dt)$ in the region. In virtue
of (\ref{eqc1})

\begin{eqnarray}
p_\tau (t) = \lambda f(\lambda t).
\nonumber
\end{eqnarray}
Equation (\ref{eqc14}) and the initial conditions for
$P\{N(0,t)=0\} =u(t)$  having the
form $u(0)=1$ and $u'(0)=-\lambda$  specify $u(t)$ uniquely and yield
$u(t)=\varphi (\lambda t)$, where

\begin{eqnarray}
\varphi (t) = 1 - t + \int\limits_0\limits^t(t-s)f(s)\,ds
\label{eqc15}
\end{eqnarray}

Since $N_1(dt)$ and $N_2(dt)$ are independent,

\begin{eqnarray}
P(N(0,t)=0) = P(N_1(0,t)=0) P(N_2(0,t)=0),
\nonumber
\end{eqnarray}
where $N=N_1+N_2$ is the sequence of events for $G_1\cup G_2$.
It follows that for any $t>0$ one has

\begin{eqnarray}
\varphi ((\lambda_1+\lambda_2)t) = \varphi (\lambda_1 t) \varphi (\lambda_2 t).
\nonumber
\end{eqnarray}
or

\begin{eqnarray}
\psi (t) = \psi (pt) + \psi (qt)
\label{eqc16}
\end{eqnarray}
where $\psi (t)=\ln \varphi (t)$, \,$p=\lambda_1/(\lambda_1+\lambda_2)$
and $p+q=1$. Iteration of (16) yields for $p=q=1/2$

\begin{eqnarray}
\psi (t) = \psi (\varepsilon_n t)/\varepsilon_n,
\quad \varepsilon_n = 2^{-n}
\nonumber
\end{eqnarray}
or

\begin{eqnarray}
\frac {\psi (t)}{\psi (1)} =
\frac {\psi (\varepsilon_n t)}{\psi (\varepsilon_n)}.
\label{eqc17}
\end{eqnarray}
By $\varphi (0)=1$, \,$\varphi' (0)=-1$ we have
$\psi (0)=0$ and $\psi'(0)=-1$. Using L'Hospital's rule we will have

\begin{eqnarray}
\lim_{n\to \infty}
\frac {\psi (\varepsilon_n t)}{\psi (\varepsilon_n)} =
\lim_{n\to \infty}
\frac {\psi' (\varepsilon_n t)t}{\psi' (\varepsilon_n)} = t.
\nonumber
\end{eqnarray}
By (\ref{eqc17}) one has $\psi (t)=-\alpha t$ or
$\varphi (t)=\exp (-\alpha t)$.
However, in that case
$f(x)=\alpha^2e^{-\alpha x}$ and $\int f(x)\,dx=1$,
whence $\alpha =1$. Statement 4 is proven for $p=q=1/2$.

In the general case, $p\ne 1/2$, the iteration of (\ref{eqc16}) yields

\begin{eqnarray}
\psi (t) = \sum\limits^n\limits_{k=0}C_n^k\psi (\varepsilon_{k,n}t)
\nonumber
\end{eqnarray}
where $\varepsilon_{k,n}=p^kq^{n-k}$.

As above one has

\begin{eqnarray}
\psi (\varepsilon_{k,n}t) = \psi (\varepsilon_{k,n}) (t+\delta_{kn})
\nonumber
\end{eqnarray}
with $\delta_{kn}=o(1)$ as $n\to \infty$.

Using (\ref{eqc15}) and the apriori bound

\begin{eqnarray}
f(x) < cx^{-\theta},\quad 0<\theta <1,\quad 0 < x < \varepsilon
\nonumber
\end{eqnarray}
it is easy to show that

\begin{eqnarray}
\vert \delta_{k,n}\vert < k_t \cdot [\max (p,q)]^{n(1-\theta)}.
\nonumber
\end{eqnarray}
Therefore we have again $\psi (t)=\alpha t$ because $\delta_{k,n}$
are small uniformly in $k$.
The proof of the statement 4 is complete.

\bigskip
\bigskip

\section{Conclusion}

We have presented a theoretical analysis of the distribution of
interevent interval $\tau$ in a point process. It is shown that,
when assumptions natural to seismic events have been made, the
distribution of $\tau$ may be a function of a single parameter,
the rate $\lambda$, provided the distribution is exponential. This
contradiction means that the nature of the empirical unified
distribution of $\tau$ is more complicated.
One has to sepatate universal properties of $\tau$ from a visual
artifical effects.

Making very general assumptions, we have found how the
distribution of $\tau$ behaves near 0 and $\infty$. As was to be
expected \cite{Bak}, the behavior is related to that of the Omori law
near zero and to the Poisson character of main seismic events,
when one deals with asymptotics at infinity. It is these
asymptotics which essentially make the probability density for
$\tau$ "universal" in \cite{Bak}, when plotted on a log-log scale.

The parameterization of the distribution of $\tau$ put
forward in \cite{Corr} for $\lambda t>0.05$  has the form
$f(x)=cx^{\gamma -1}\exp(-x/a)$. It was shown
above that the parameter $1/a$ can be treated as the fraction of
main events among all seismic events. The estimate $a=1.23$
derived in \cite{Corr} yields $a^{-1}\approx 80\%$, which can hardly
be a universal constant. The main events in Italy are $60\%$ among
the $m\ge 3.5$ events (see \cite{Mol}).

The factor $x^{\gamma -1}$ is missing in the formula for f in the
models considered above. This factor may be replaced (see (\ref{eqc7}))
by a factor of the type $\exp (-cx^{1-\theta})$, if the aftershock
rate decays as a power function $t^{-1-\theta}$,
$0<\theta \le 1$; the factor degenerates to a
constant for $\theta >1$. Consequently, it remains an open question
as to what is the physical meaning of $\gamma$.


\newpage
{\bf Appendix}
\bigskip

{\it Proof of Statement 1}.

We are going to find the asymptotics of $P(\tau >t)$ as $t\to \infty$
using (\ref{eqc6}). To do this, the following three limits should be found as
$t\to \infty$:

\begin{eqnarray}
A: = P(N(0,t)=0) \to P(N_0(0,\infty)=0) = P(\tau_{cl}=0).
\nonumber
\end{eqnarray}
Here, $\tau_{cl}$ is the cluster duration in $N_0$.

\begin{eqnarray}
B: = \int\limits_0\limits^\infty P(N_0(du)>0,\,N_0(u,u+t)=0) \to
\int\limits_0\limits^\infty P(\tau_{cl}\in du) = P(\tau_{cl}>0).
\nonumber
\end{eqnarray}
Consequently, $A+B\to 1$. It remains to find the limit for the
expression under the $\exp$ sign in (\ref{eqc6}). One has

\begin{eqnarray}
\nonumber
C: &=&\int\limits_0\limits^\infty P(N_0(u,t+u)>0)\,du \to
\int\limits_0\limits^\infty P(N_0(u,\infty)>0)\,du = \\
\nonumber
   &=&E\,\int\limits_0\limits^\infty {\bf 1}_{N_0(u,\infty)>0}\,du
    = E\, \int\limits_0\limits^{\tau_{cl}}du = E\tau_{cl}.
\nonumber
\end{eqnarray}
We have used the notation ${\bf 1}_A : {\bf 1}_A=1$, if $A$ is true
and ${\bf 1}_A=0$ otherwise.
It remains to substitute the resulting limits into

\begin{eqnarray}
P(\tau >t) = \exp \{-\lambda^*(t+C)\} [A+B]/(1+\Lambda).
\label{eqc18}
\end{eqnarray}
We now are going to prove the second part of Statement 1.
Let $E\tau_{cl}=\infty$. The asymptotics of $C$ then calls for
refinement. One has

\begin{eqnarray}
P(N_0(u,t+u)>0) \le E\,N_0(u,t+u) =
\int\limits_u\limits^{t+u}\lambda_0(v)\,dv.
\nonumber
\end{eqnarray}
If $\lambda_0(v)=c\theta(1-\theta)v^{-(1+\theta)}$ for $v\gg 1$, then

\begin{eqnarray}
C < \int\limits_0\limits^\infty \,du
\int\limits_u\limits^{t+u}\lambda_0(v)\,dv = ct^{1-\theta}(1+o(1)),
\quad t\to \infty,
\nonumber
\end{eqnarray}
as follows from L'Hospital's rule. Relation (\ref{eqc7}) stands proven.

{\it Proof of Statement 2}.

The distribution density for $\tau$ can be found by
differentiating (\ref{eqc18}). To do this, we make differences for the
functions $A$, $B$, $C$ in (\ref{eqc18}). One has for small $\delta$:

\begin{eqnarray}
\nonumber
[A(t)-A(t+\delta )]/\delta &=&P(N(t,t+\delta)>0)\,\delta^{-1}
(1-P(N_0(0,t)>0\,\vert \,N_0\{ t\} =1) \\
\nonumber
                            &\simeq &\lambda_0(t)(1+o(1)).
\nonumber
\end{eqnarray}
The last conclusion follows from the regularity requirement
imposed on (\ref{eqc8}). Similarly, one has for $B(t)$ using (\ref{eqc9}):

\begin{eqnarray}
\nonumber
[B(t)-B(t+\delta )]/\delta =
\int\limits_0\limits^\infty [P(N_0(du)>0,\,N_0(u+t,u+t+\delta)>0)- \\
\nonumber
-P(N_0(du)>0,\,N_0(u,u+t)>0,\,N_0(u+t,u+t+\delta)>0)]= \\
\nonumber
= \int\limits_0\limits^\infty \lambda_0(u)\lambda_u(t)\,du (1+o(1)),
\qquad \qquad \qquad \quad \quad \, \,
\nonumber
\end{eqnarray}
where
$\lambda_u(t)=P(N_0(u+t+\delta)-N_0(u+t)>0\,\vert \,N_0\{ u\} =1)/\delta$.
One has $C'(t)=-\Lambda(1+o(1))$ for $C(t)$ using (\ref{eqc8}).
It remains to differentiate (\ref{eqc18}) and then to substitute
the resulting asymptotic
expressions for the derivatives $A'$, $B'$, $C'$ and the values
$A(0)=1$, $B(0)=\Lambda$ and $C(0)=0$.

{\it Proof of Statement 3}.

It follows from the description of the cascade generation of
$N_0$ that its rate $\lambda_0(t)$ satisfies the integral equation

\begin{eqnarray}
\lambda_0(t) = \int\limits_0\limits^t \pi_0(x)\,
\lambda_0(t-x)\,dx + \pi_0(t),
\label{eqc19}
\end{eqnarray}
where $\pi_0(t)$ is the rate of rank 1 events. Iteration of (\ref{eqc19})
then yields

\begin{eqnarray}
\lambda_0(t) = \pi_0(t) +\pi_0 *\pi_0 (t) + \pi_0 *\pi_0 *\pi_0 (t) + \ldots
\nonumber
\end{eqnarray}
If one passes to the Laplace transform, $\lambda \to \widehat{\lambda}$,
then both relations for $\lambda_0(t)$ are reduced to the form

\begin{eqnarray}
\widehat{\lambda}_0(s) = \widehat{\pi}(s)/(1-\widehat{\pi}(s)).
\nonumber
\end{eqnarray}

Let $\pi (t)$ be monotone near 0 and $\infty$. Assume also
that $\pi (t)$ behaves like a power law:
$\pi_0(t)\sim c_0t^{-p}$, $t\ll 1$ or
$\pi_0(t)\sim c_1t^{-1-\theta}$, $t\gg 1$, where
$0<p,\theta <1$. In that case the use of the Tauberian theorems
(see \cite{Fel},
Ch. 13 and Ch. 17, \S 12) yields conclusions of the form
$\lambda_0(t)/\pi(t)\to \mbox {\rm const}$
as $t\to 0$ or $t\to \infty$, respectively.

We now are going to prove (\ref{eqc13}). Consider the rate of a pair
of events in an $N_0$ cluster:
$\lambda_0(u,v) = P(N_0(du)=1, N_0(dv)=1) /(du\,dv),\, \,u<v$.
Recalling that this is a
cascade generation of $N_0$, the states $u$ and $v$ in $N_0$ can
be derived in two ways. The one is when $u$ and $v$ have no
common parent except $t=0$; the second is when $u$ and $v$ have a
common parent $z$ in the first generation (a state of rank 1). If the
common parent $z$ for u and v has rank $r>1$, then the probability
of that event will be of order $O((dz)^2\,du\,dv)$, which is negligibly small
compared with $O(dz\,du\,dv)$. This consideration leads to the
following equation for $\lambda_0(u,v)$:

\begin{eqnarray}
\lambda_0(u,v) = \lambda_0(u)\lambda_0(v) + \int\limits_0\limits^u
\pi_0(z)\lambda_0(u-z, v-z)\,dz,\quad u<v.
\label{eqc20}
\end{eqnarray}
Put  $a_t(u)=\lambda_0(u,u+t)$,
\quad $b_t(u)=\lambda_0(u)\lambda_0(u+t)$. Then (\ref{eqc20}) gives

\begin{eqnarray}
a_t(u) = b_t(u) + \pi_0 * a_t(u).
\nonumber
\end{eqnarray}
Whence

\begin{eqnarray}
a_t(u) &=& b_t(u) +
b_t(u) * (\pi_0 +\pi_0 *\pi_0 + \pi_0 *\pi_0 *\pi_0 + \ldots) = \nonumber\\
&=& b_t(u) + b_t(u) * \lambda_0(u).
\label{eqc21}
\end{eqnarray}
We are interested in the conditional rate in a $N_0$
cluster:

\begin{eqnarray*}
\lambda_u(t) = \lambda_0(u,u+t)/\lambda_0(u).
\end{eqnarray*}
One has from (\ref{eqc21}) using the notation $a_t$ and $b_t$:

\begin{eqnarray*}
\lambda_u(t) = \lambda_0(u+t) +
\int\limits_0\limits^u\lambda_0(x) \lambda_0(x+t)\lambda_0(u-x)\,dx
/\lambda_0(u).
\end{eqnarray*}
It remains to substitute that expression in (\ref{eqc10}). One has

\begin{eqnarray*}
p_\tau(t) = [\lambda_0(t) &+&
\int\limits_0\limits^\infty\lambda_0(u) \lambda_0(u+t)\,du \cdot
(1+\Lambda) + \\
&+& \lambda (1+\Lambda)] / (1+\Lambda)\cdot(1+o(1)),
\quad t\to 0.
\end{eqnarray*}
However, $(1+\Lambda)^{-1}=1-\lambda_{\pi}$, so (\ref{eqc13}) is proved.

In order to have
$p_\tau(t)=O(\lambda_0(t))$  as $t\to 0$ , one has to show that
$\lambda_0(u+t)<k\lambda_0(t)$ for small $t$. To do this, we demand
$\pi_0(t)=c_1t^{-\theta},\, 0<t<\varepsilon$ and $\pi_0(t)<\varphi(t)$.
Here, $\varphi$ is a smooth function,
$\int\limits_0\limits^\infty \varphi(t)<1$ and
$\varphi(t) \sim ct^{-1-\theta}$, $t\gg 1$ with $0<\theta <1$; also,
$\varphi =\pi_0$ for $t<\varepsilon$. Then

\begin{eqnarray*}
\lambda_\varphi = \varphi + \varphi * \varphi +
\varphi * \varphi * \varphi \ldots
\end{eqnarray*}
is a smooth function. One has $\lambda_0(t)<\lambda_\varphi(t)$
in virtue of (\ref{eqc12}), since
$\pi_0\le \varphi$. One has $\lambda_\varphi(t)/\varphi(t) \to c$
as $t\to \infty$ from the power law behavior
of $\varphi$ at $\infty$ (see Statement 3a). One also has
$\lambda_0(t)/\pi_0(t) \to 1$ as
$t\to 0$, hence $\lambda_0(t)\uparrow \infty$ as $t\to 0$.
Consequently, $\max\limits_{t>t_0}\lambda_0(t)$ will
coincide with $\lambda_\varphi(t_0)=\lambda_0(t_0)(1+o(1))$
starting from some small $t_0$. Hence

\begin{eqnarray*}
\lambda_0(u+t) \le \max\limits_{v>t_0}\lambda_0(v) \le
\max\limits_{v>t_0}\lambda_\varphi(v) \simeq \lambda_0(t_0)(1+o(1)).
\end{eqnarray*}


\begin{thebibliography}{ABCDA}


\bibitem[1]{Bak}
Bak, P., Christensen, K., Danon, L., and Scanlon, T. Unified
scaling law for earthquakes. {\it Phys. Rev. Lett}. {\bf 88}, 178501, 2002.

\bibitem[2]{Cor}
Corral, A. Local distributions and rate fluctuations in a
unified scaling law for earthquakes. {\it Phys. Rev. E}, {\bf 68},
035102(R), 2003.

\bibitem[3]{Corr}
Corral, A. Time-increasing hazard and increasing time until
the next earthquake. {\it ArXiv: cond-mat}/0310407 v1, 170ct 2003.

\bibitem[4]{Uts}
Utsu, T., Ogata, Y., and Matsu'ura, R.S. The centenary of the
Omori formula for a decay law of aftershock activity. {\it J. Phys.
Earth} {\bf 43}, 1-30, 1994.

\bibitem[5]{Dal}
Daley, D.J., and Vere-Jones, D. {\it An Introduction to the
Theory of the Point Processes}. N.-Y., Berlin: Springer-Verlag,
1988, 702pp.

\bibitem[6]{Ver}
Vere-Jones, D. Stochastic models for earthquake occurrence.
{\it J. Roy. Statist. Soc}. {\bf B32}, 1-62, 1970.

\bibitem[7]{Haw}
Hawkes, A.G., and Adamopoulos, L. Cluster models for
earthquakes - regional comparisons. {\it Bull. Int. Stat. Inst.},
{\bf 45}: 3, 454-461, 1973.

\bibitem[8]{Oga}
Ogata, Y. Statistical models for earthquake occurrences and
residual analysis for point processes. {\it Mathematical Seismology}
{\bf 1}, 228-281, Inst. Statist. Math., 1986.

\bibitem[9]{Sai}
Saichev, A., and Sornette, D. Anomalous power law
distribution of total lifetimes of aftershock sequences.
{\it ArXiv: physics}/0404019 v1, 4 Apr 2004.

\bibitem[10]{Mol}
Molchan, G., Kronrod, T., Dmitrieva, O., and Nekrasova,
A. Seismic risk oriented multiscale seismicity model: Italy.
{\it Computational Seismology} 28 (in Russian), p.193-224, 1996.

\bibitem[11]{Fel}
Feller, W. {\it An Introduction to Probability Theory and Its
Applications} II. N.-Y., John Wiley and Sons, Inc., 1966, 740pp.

\end{thebibliography}
\end{document}